\documentstyle[12pt]{article} 
\def \be{\begin{equation}} 
\def \ee{\end{equation}} 
\begin{document} \begin{flushright} TIFR/TH/97-16\\ 
April 29, 1997
\end{flushright} 
% The following expanded version is dated 5 September 1997.
\bibliographystyle{unsrt} \vskip0.5cm
\baselineskip=.8cm \begin{center} { 
\LARGE 
\bf CP-violating correlations between lepton-charge and jet-charge 
in $Z\rightarrow b\bar b$ events}\\ 
[7mm] { 
\bf G.V. Dass$^a$ and
K.V.L. Sarma}$^{b,*}$\\ [3mm] 
$^a${ \it Department of Physics, Indian
Institute of Technology, Powai, Mumbai, 400 076, India }\\ \vskip0.3cm
 $^b${ \it Tata Institute of Fundamental Research, Homi Bhabha Road,
Mumbai, 400 005, India } \\ [10mm] 
\end{center}

\bigskip 

\begin{center} Abstract \end{center} 
\bigskip

\baselineskip=.8cm 

Correlations that signal direct $CP$ violation in the bottom sector
are suggested. They need data on $Z\rightarrow b\bar b$ events which
consist of a bottom hadron decay into an exclusive semileptonic
channel $h \ell \nu $ (where $h$ is a hadron), and the other bottom
hadron decay into a multi-body final state that manifests as a jet in
the opposite hemisphere.

\bigskip

\bigskip

\noindent PACS: 13.20.He, 11.30.Er.\\ {
\it {Keywords:}} Bottom decays,
Correlations, Jet charge, CP violation, Oscillations.

\vfill

$^*$~E-mail: kvls@theory.tifr.res.in;~~fax: 091 22 215 2110 

\newpage 
\baselineskip=.8cm 

The question whether $CP$ symmetry is violated in a sector other than
the neutral kaons is of considerable interest in particle physics. In
the bottom sector, present data give only limits on $CP$ violation in
the mixing of neutral $B$ mesons. The limit coming from the $\Upsilon
$(4S) decay \cite{CLEO} is based on data on the charge asymmetry
between likesign dilepton events. At the $p\bar p$ collider \cite{CD},
asymmetry between ($\mu ^+\mu ^+$) and ($ \mu ^-\mu ^-$) events is
used to constrain $CP$ violation in mixings in the $B^0$ and $B_s^0$
systems. Theoretical estimates based on the standard model indicate
rather tiny values ($|\epsilon |\sim 10^{-3}$) for mass-matrix $CP$
violation in the $b$ sector \cite{{ALT},{HAG}}; they are smaller than
the current experimental upper limits by an order of magnitude.  On
the other hand, effects due to direct $CP$ violation are expected to
be sizeable in $B$ decay; they await observation at the upcoming
asymmetric $B$ factories. Here we wish to point out that it may be
possible to obtain signals for direct $CP$ violation in $b$ decays by
measuring certain charge correlations in $Z\rightarrow b\bar b$
events.

We consider the time-distribution of a bottom hadron decay into some
specific channel (for instance, $B^0\rightarrow D^{*\mp }\ell ^{\pm }
\nu _{\ell }$), while the flavour of this hadron at production is
determined by the `jet charge' of the $b$-jet in the opposite-side
hemisphere. Such measurements have already been made by various groups
(\cite{WU}-\cite{SLD}) in demonstrating the phenomenon of $B^0\bar B^0$
oscillations. From the data on events consisting of $B$ semileptonic
decay and the opposite-side jet $J^{\pm }$ having jet-charge $\pm
$(1/3), we envisage asymmetries involving the charge combinations $(
J^- \ell ^+\, - \,J^+\ell ^-)$ and $( J^- \ell ^- \, - \,J^+\ell ^+)$;
such asymmetries serve as signals for $CP$ violation in the
$b$-sector.

To put the present proposal in perspective, we turn to the 
experimental efforts made in obtaining the frequency $\Delta m $ of
$B^0 \bar B^0$ oscillations from $Z \rightarrow b\bar b$ decays. One
type of experiments \cite{ll} deals with the combinations $(\ell
^+\ell ^+ ~+~ \ell ^-\ell ^- )$ and $( \ell ^+\ell ^- ~+~ \ell ^-\ell
^+)$ which are $CP$-even. The corresponding $CP$-odd combinations
$(\ell ^+\ell ^{\pm } ~-~ \ell ^-\ell ^{\mp })$ are normally not
considered; the relevant $CP$ violation, being of the mass-matrix
variety, is presumably small. In the lepton-jet method, the $CP$-even
combinations $(J^-\ell ^+ ~+~ J^+\ell ^-)$ and $( J^-\ell ^- ~+~
J^+\ell ^+)$ have been studied by the ALEPH \cite{AL96}, OPAL
\cite{OP} and DELPHI \cite{ll} groups. The corresponding $CP$-odd
combinations $(J^-\ell ^+ ~-~ J^+\ell ^-)$ and $( J^-\ell ^- ~-~
J^+\ell ^+)$ motivate the present suggestion. Although the mass-matrix
$CP$ violation can occur at both ends, our interest is focussed on the
direct $CP$ violation arising at the jet-end.

{\bf Production and Decay Tags:} Let the $Z$ decay produce 
a $b$-hadron
which in turn decays by producing a particle jet; let the jet-charge
using the weighted sum of the individual track charges (for details,
see, e.g., \cite{OPAL}) be measured. We denote the rate of production
of the $b$-jet with normal jet charge (=$-$1/3) by $P_n$. The jet from
a b-quark can also have the abnormal charge (=+1/3) if the $b$
fragmented into $\bar B^0$ or $\bar B_s^0$ which oscillated to the
conjugate meson having positive bottom flavour; let $P_a$ denote the
production rate of the $b$-jet with abnormal jet-charge. The rates
associated with jet production then are
\begin{eqnarray} 
P_n &:& (b \rightarrow J^-),~~P_a~:~ (b \rightarrow
J^{+}), \\ \bar {P_n} &:& (\bar b \rightarrow
J^{+}),~~\bar {P_a}~:~ (\bar b \rightarrow J^{-}); 
\end{eqnarray}
here the superscript ($\pm $) on $J$ refers to the sign of the jet
charge $\pm $(1/3). For later use, we define the fractional
differences 
\be 
\pi _n\equiv {P_n-\bar P_n \over P_n+\bar P_n }~,~~
\pi _a\equiv { P_a-\bar P_a\over P_a+\bar P_a}~.
                                             \label{pi} 
\ee 
If $CP$ invariance were valid, we have $\pi _n = 0 = \pi _a$.

As for the decay tag, we focus on the $B^0$ meson undergoing
semileptonic decays. The case of $B^0$ is instructive as it involves
the complications arising from mixing, and has a large body of data
available; the cases of other hadrons like $B^+$ and $\bar \Lambda _b$
are simpler. Among the semileptonic decays, we choose a channel that
contains {\it a single hadron}, $ B^0 \rightarrow h\, \ell \,\nu$. For
the decay into such a channel, $CPT$ invariance forces the amplitude
$A$ for $B^0$ decay and the amplitude $\bar A$ for $\bar B^0$ decay to
be complex conjugates of each other, apart from a negligible
correction from electroweak phase $(\bar A= A^*\, \exp~[2i\delta
_{ew}])$; this assumes that the final state polarizations are not
measured and second order weak effects are ignorable. Hence for decays
into such channels, one gets 
\be |\bar A|=|A|~. \label{cpt} \ee 
For definiteness, we shall focus on the decay $B^0 \rightarrow
D^*(2010)\,\ell \nu $.
 
Let $t=0$ be the instant of production of the $b\bar b$ pair (we
ignore events in which $Z$ decays into two or more $b\bar b$
pairs) and $t$ be the time when the $B^0$ decays. We define the decay
rates into normal ($D_n$) and abnormal ($D_a$) modes by
\be D_n(t) = \Gamma
(\bar B^0(t) \rightarrow D^{*+}\, \ell^- \,\bar {\nu _{\ell }}),~~
D_a(t) = \Gamma (\bar B^0(t) \rightarrow D^{*-}\, \ell^+\, {\nu
_{\ell }}); \label{ds} \ee
\be \bar D_n(t) = \Gamma (B^0(t) \rightarrow D^{*-}~ \ell^+~{\nu
_{\ell } }),~~ \bar D_a(t) = \Gamma (B^0(t) \rightarrow D^{*+}~
\ell^-~ \bar {\nu _{\ell } }). \label{dbs}  
\ee 
Here $\bar B^0(t) $ denotes the physical state that evolved upto
time $t$ from an initial $\bar B^0$ state; the probabilities for an
initial state corresponding to an antiquark ($\bar b$) have a `bar' on
them. Clearly the abnormal decay probabilities $D_a$ and $\bar D_a$
arise due to $B^0\bar B^0$ oscillations.

Differences between particle and antiparticle decay rates (ignoring
an over-all factor which is time-independent) are given by
\begin{eqnarray} 
D_n(t) -\bar D_n(t) & =&(\,|\bar A|^2-|A|^2\,)\, \exp
(-\Gamma t)\, \cos^2 {\Delta m \over 2} \,t~\\ 
&=& 0\,\, ; 
                    \label{ddn}
\\ 
D_a(t)-\bar D_a(t) & =& (\,|{p\over q}A|^2 -|{q\over
p}\bar A|^2\, )\, \exp (-\Gamma t)\, \sin^2 { \Delta m \over 2}\,t~\\
&\simeq & 0.\,\, 
                       \label{dda} 
\end{eqnarray} 
The notation is standard: the parameters $p$ and $q$ define $CP$
violation in mixing; $\Delta m =m_2-m_1~;~ \Gamma \simeq \Gamma _1
\simeq \Gamma _2~$, ignoring any width-difference $\Delta \Gamma \simeq
0$. We used the $CPT$ invariance result, $|\bar A |=|A| $. Equation
(\ref{dda}) further assumes that $CP$ violation arising from mixing is
negligible; this may be reasonable since theoretical estimates
\cite{{ALT},{HAG}} show that $|q/p|\simeq 1+{\cal O}(10^{-3}) $. For
decays of $b$-hadrons which do not oscillate, eq. (\ref{dda}) is exact
since $\Delta m$ is to be set zero.

Next we write the relative numbers of events $N^{ji}$ wherein the
first superscript $j$ is the sign of the jet charge and the 
second superscript $i$ is the sign of the lepton charge in the decay
$B^0\rightarrow h \,\ell ^i \,\nu \,$: 
\begin{eqnarray} 
N^{-+} &=& P_n\,\bar D_n + \bar P_a\, D_a~,    \label{enf}\\ 
N^{+-} &=& \bar P_n\, D_n + P_a\, \bar D_a~,\\ 
N^{++} &=& \bar P_n\, D_a + P_a \,\bar D_n~, \\ 
N^{--} &=& P_n\, \bar D_a + \bar P_a \,D_n~.     \label{enl}  
\end{eqnarray} 
These expressions are easy to write down: for instance, $N^{-+}$
refers to events with negative jet charge arising from the decay of a
$b$-hadron which did not oscillate (termed normal), or from the decay
of a $\bar b$-hadron which did oscillate (termed abnormal); the
corresponding positively-charged lepton can result from the decay
following either the transition $B^0(t)\rightarrow B^0$ or $\bar
B^0(t)\rightarrow B^0$.

We rewrite eqs. (\ref{enf}-\ref{enl}) in terms of the $CP$-violating
differences $\pi _n$ and $\pi _a$ by using eqs. (\ref{ddn}) and
(\ref{dda}), as
\begin{eqnarray} 
N^{-+} &=& p_n\, d_n\, (1+\pi _n) +
p_a \,d_a\, (1-\pi_a)~,\\ 
N^{+-} &=& p_n\, d_n\, (1-\pi _n) + p_a\, d_a\, (1+\pi
_a) ~, \\ 
N^{++} &=& p_n\, d_a \,(1-\pi _n) + p_a \,d_n \,(1+\pi _a) ~, \\
N^{--} &=& p_n\, d_a\, (1+\pi _n) + p_a\, d_n \,(1-\pi _a) ~, 
\end{eqnarray}
wherein $p$'s and $d$'s denote the averages 
\begin{eqnarray}
p_i &\equiv & {P_i+\bar P_i\over 2} \,,\,\,\, (i=n,a\,),\\ 
d_i &\equiv & {D_i+\bar D_i\over 2} \,,\,\,\, (i=n,a \,).
\end{eqnarray}

{\bf $CP$-violating Correlations:} We now construct a 
$CP$-violating correlation between jet-charge and lepton-charge:
\begin{eqnarray} 
\delta _{J\ell }(t) &= & 
{ N^{-+}- N^{+-}\over  N^{-+}+ N^{+-} } 
                                             \label{d1} \\
 &=&{p_n\, d_n\, \pi _n- p_a \,d_a \,\pi _a 
\over p_n\,d_n+p_a\,d_a }~.                   \label{d2}
\end{eqnarray}
Upon substituting the relation $(d_a/d_n)=\tan ^2({1\over 2}\Delta
mt)$ and defining $\xi =(p_a/p_n)$, we get
\be 
\delta _{J\ell }(t) = { \pi _n - \xi \, \pi _a \,
\tan ^2({1\over 2}\Delta mt ) \over 1+ \xi \,
                    \tan ^2({1\over 2}\Delta mt )}~.
                                                  \label{d3}
\ee
From the time dependence of this asymmetry it is easy to determine the
$CP$-violation parameters $\pi _n$ and $\pi _a$; for instance, one may
use
\begin{eqnarray}
\delta _{J\ell }(t) 
&=& +\pi _n ~,~{\rm for}~ t=2k~(\pi /\Delta m)\,
                                    , \label{d4} \\
\delta _{J\ell }(t) 
&=& -\pi _a~,~{\rm for}~ t=(2k+1)~(\pi / \Delta m)\, , 
\end{eqnarray} 
where $k =$ 0,1,2,...~.
The decay-times corresponding to $k= 0$ and 1 are in the
convenient range as $(\pi /\Delta m)\simeq $ 6 ps, which is about 4
lifetimes of the beon. Fits to the shape of the time distribution can
determine also $\xi $ and $\Delta m$. 

The time-integrated version of eq. (\ref{d3}) fixes the parameter 
combination
\be
\Delta _{J\ell }= {(2+x^2) \pi _n - x^2\xi \pi _a \over (2+x^2)
+ x^2\xi  }, \label{Del}
\ee
where the mixing parameter $x$ is given by $x= (\Delta m/\Gamma )$.
Since the average value of $x$ is $ 0.73\pm 0.05$ \cite{PDG}, the
first terms in the numerator and denominator would dominate as the
ratio $[(2+x^2)/x^2]$ is about $\sim 5$, and the production ratio $\xi
=(p_a/p_n) $ itself would not be large (as $p_a$ gets contributions
only from $B^0$ and $B^0_s$, while $p_n$ gets contributions from all
$b$-hadrons). Hence, it is reasonable to expect 
\be \Delta _{J\ell }\simeq \pi _{n}~; \ee 
in other words, $\Delta _{J\ell }$ will be a measure of the $CP$
violation due to unequal production of jets arising from $b$-hadrons
and $\bar b$-hadrons.

A minor variant of the $\delta _{J\ell }$ is the $CP$-violating
correlation which involves mixed events:
\begin{eqnarray} 
\tilde {\delta }_{J\ell }(t) 
              &= & { N^{--}- N^{++}\over  N^{--}+ N^{++} }
                                               \label{dt1} \\
&=& {p_n\, d_a\, \pi _n- p_a \,d_n \,\pi _a 
       \over  p_n\,d_a+p_a\,d_n }             \label{dt2}  \\ 
&=& {\pi _n\,\tan ^2({\Delta m \over 2}t) - \xi \, \pi _a 
\over \tan ^2({\Delta m\over 2}t ) +\xi }~.     \label{dt3} 
\end{eqnarray} 
Perhaps a convenient way to obtain $\pi _n $ and $\pi _a$ is to
measure both $\tilde \delta $ and $\delta $ at the time $t=(\pi
/\Delta m)$. We do not pursue correlations of the hybrid type
($e.g.,~[N^{ii} - N^{ij}]$) as they are not purely $CP$ violating.

{\bf Other Decay Tags:} In an effort to enlarge the data sample, it
may be useful to replace the exclusive tag $ B^0 \rightarrow h\, \ell
\,\nu$ by an inclusive lepton decay of bottom hadrons. The $CPT$
relation $|\bar A|=|A|$ of eq. (\ref{cpt}) will then be replaced by
the corresponding inclusive rate equalities, and the asymmetries
$\delta $ and $\tilde \delta $ will be given by eqs. (\ref{d2}) and
(\ref{dt2}), respectively. But, the parameters $d_i$ would not have
simple interpretation; for example, $d_a$ would get contributions
(weighted with production-fractions) from both $B^0$ and $B_s^0$ ,
while $d_n$ would get contributions also from the non-oscillatory
bottom hadrons. Consequently the simple time-dependence of the
formulas (\ref{d3}) and (\ref{dt3}) would no longer hold.
Nevertheless, the important feature that ${\delta }_{J\ell }$ and
$\tilde {\delta }_{J\ell }$ represent signals for $CP$ violation would
remain unchanged.

For bottom decays, either into semileptonic modes that have two or
more hadrons ($e.g.,~B^0 \rightarrow \mu ^+\,\nu_{\mu }\, D_s^-\,
K^0$), or into purely hadronic channels ($e.g.,~ B^+\rightarrow \pi
^+\, \pi ^+\,D^- $), the $CPT$ invariance condition of eq. (\ref{cpt})
need not hold. However for any general decay, we define, as in
eqs. (\ref{ds}) and (\ref{dbs}), the four decay rates $D_i$ and $\bar
D_i$ where $i=(n,a)$; in the case of a non-oscillating bottom hadron
one needs to set $D_a=\bar D_a=0$. The $CP$-violating parameters,
analogous to $\pi _{n,a} $ of eq. (\ref{pi}), are now definable at the
decay-end also:
\be 
\Delta _n\equiv {D_n-\bar D_n \over D_n+\bar D_n }~,~~
\Delta _a\equiv { D_a-\bar D_a\over D_a+\bar D_a}~.  
\label{dee} 
\ee 
Thus, for any specific decay channel indicated by `tag' $T$ (and the
corresponding antiparticle channel by $\bar T$), the correlations of
eqs. (\ref{d1}) and (\ref{dt1}) are generalized to give
\be 
\delta _{JT}(t) 
= { (\pi _n-\Delta _n) - \xi\, (\pi _a - \Delta _a)\,
                             \tan ^2({\Delta  m\over 2}t ) 
               \over 
             (1- \pi _n\Delta _n) + \xi \, (1-\pi _a \Delta _a)  
                    \tan ^2({\Delta m\over 2}t )}~, \ee
\be
\tilde \delta _{JT}(t) 
= { (\pi _n-\Delta _a)\tan ^2({\Delta  m\over 2}t) - 
                          \xi\, (\pi _a - \Delta _n)\,
               \over 
             (1- \pi _n\Delta _a) \tan ^2({\Delta  m\over 2}t )
                    + \xi \, (1-\pi _a \Delta _n) } ~. 
\ee 
The formulas become especially simple when the tag refers to the decay
of a $b$-hadron that does not oscillate: we set $\Delta m =0$, neglect
second-order $CP $ violations ($\pi _i\Delta _n\simeq 0$), and obtain
the relations
\begin{eqnarray}
\delta _{JT}&=& (~\pi _n -\Delta _n),\\
\tilde \delta _{JT}&=& (-\pi _a+\Delta _n),
\end{eqnarray}
which are independent of decay time.

{\bf Discussion:} The program envisaged in this note is in some sense
complementary to that at the $B$ factories. It needs $Z$ decays giving
$b$-jets which are due to multi-body decays of $b$-hadrons. In
contrast, the $\Upsilon $(4S) gives only non-relativistic $B^0\bar B^0
$ and $B^+B^-$ pairs which cannot lead to jets. In fact, at the $B$
factories, special two-body channels (e.g., $J/\psi ~K_S^0~$) will be
of primary interest for testing the Standard Model mechanism for $CP $
violation.

In $Z$ events, we consider decays of bottom hadron $h_b$ giving rise
to jets $f_J$ with specified jet-charge and construct the rate
difference: \be \Gamma (h_b \rightarrow f_J) - \Gamma (\bar h_{\bar b}
\rightarrow \bar f_J).  \ee Although this does not vanish by $CPT$
invariance (as the channels which do not give rise to jets are
excluded), it is of little practical interest since it requires the
particle originating the jet to be identified as $h_b$. Hence we sum
over the initial bottom hadrons. Denoting by $F$ the fragmentation
rate of the bottom quark into the hadron, we consider 
\be
\sum_{h_b}~F(b\rightarrow h_b)~\Gamma (~h_b\rightarrow f_J)~-~
\sum_{\bar h_{\bar b}}~F(\bar b\rightarrow \bar h_{\bar b})~\Gamma
(~\bar h_{\bar b} \rightarrow \bar f_J).  
\ee 
It is this difference involving all hadrons $h_b$ and $\bar h_{\bar
b}$ decaying into jet channels $f_J$ and $\bar f_J$, that contributes
to the $CP$-violation parameters $\pi _{n,a}$. Summation over initial
state enriches the data sample, and measures the cumulative
effect of $CP$ violation originating in many parents. If the data
reveal that $\pi _a$ is vanishingly small, one may conclude, barring
conspiratorial cancellations, that large $CP$ violation is not present
in multi-body decays of the oscillatory hadrons, namely $B^0$ and
$B_s^0$.

There is yet another difference with respect to the situation at
$\Upsilon $(4S). The difference ($\ell ^+\ell ^+ - \ell ^-\ell ^-$)
between likesign dilepton rates at $\Upsilon $(4S) gives $CP$
violation in the mixing of beon states \cite{OKUN}, whereas the
correlations $\delta $ and $\tilde \delta $ refer to rate asymmetries
associated primarily with $CP$ violation of the direct kind. One
expects direct $CP$-violation effects to be larger than those due to
the state mixing, and so we dropped the latter effects.

It is difficult to estimate the expected size of the asymmetries we
have proposed. However we expect the experimental effort involved to
be roughly comparable to that in a typical $B^0\bar B^0$ oscillation
experiment. In order to establish the asymmetry $\delta $ to be
nonzero to, say, within 3 standard deviations, the number of events in
the sample needs to be $\simeq (3/\delta )^2$. Thus for the values
$\delta \simeq 0.1$ we need about $10^3$ events; such data samples are
comparable to, for instance, the $778 \pm 84$ event-sample of the
($D^*\ell ~/~Q_J $) data of the OPAL group \cite{OP}. In the
interesting case when the method used is that of the inclusive-lepton
against jet, the data sample obviously would be larger. One could
easily be dealing with samples of $\sim 10^4$-$10^5$ events, as in the
experiments of ALEPH group \cite{AL96} or DELPHI group \cite{ll},
which imply sensitivities to values of $ \delta _{J\ell }$ at the
level of a few per-cent.

In conclusion, experimental information on the correlations $\delta
_{J\ell }$ and $\tilde \delta _{J\ell }$ (or their generalized
versions $\delta _{JT}$ and $\tilde \delta _{JT }$) would provide
valuable diagnostic signals for $CP$ violation in the bottom sector.

\end{document}